\documentclass[sigconf]{acmart}
\AtBeginDocument{%
  }

\setcopyright{acmlicensed}
\copyrightyear{2018}
\acmYear{2018}
\acmDOI{XXXXXXX.XXXXXXX}
\acmConference[ASE’25]{ASE’25}{Sun 16 - Thu 20 November 2025}{Seoul, South Korea}

\acmISBN{978-1-4503-XXXX-X/2018/06}

\usepackage{color}
\usepackage{tabularray}
\usepackage{caption}        
\usepackage{tcolorbox}
\tcbset{
  colback=white,
  colframe=black,
  boxrule=0.6pt,      
  arc=0pt,           
  left=4pt,right=4pt, 
  top=3pt,bottom=3pt,
  fontupper=\footnotesize\itshape
}
\renewcommand{\refname}{\uppercase{REFERENCES}}
\begin{document}

\title{Uncovering Systematic Failures of LLMs in
Verifying Code Against Natural Language
Specifications}

\author{Haolin Jin}
\affiliation{%
  \institution{University of Sydney}
  \city{Sydney}
  \state{NSW}
  \country{Australia}}
\email{hjin3177@uni.sydney.edu.au}

\author{Huaming Chen}
\affiliation{%
  \institution{University of Sydney}
  \city{Sydney}
  \state{NSW}
  \country{Australia}}
\email{huaming.chen@sydney.edu.au}

\begin{abstract}
Large language models (LLMs) have become essential tools in software development, widely used for requirements engineering, code generation and review tasks. Software engineers often rely on LLMs to assess whether system code implementation satisfy task requirements, thereby enhancing code robustness and accuracy. However, it remains unclear whether LLMs can reliably determine whether the code complies fully with the given task descriptions, which is usually natural language specifications. 
In this paper, we uncover a systematic failure of LLMs in evaluating whether code aligns with natural language requirements. Specifically, with widely used benchmarks, we employ unified prompts to judge code correctness. Our results reveal that LLMs frequently misclassify correct code implementations as either ``not satisfying requirements'' or containing potential defects. Surprisingly, more complex prompting, especially when leveraging prompt engineering techniques involving explanations and proposed corrections, leads to higher misjudgment rate, which highlights the critical reliability issues in using LLMs as code review assistants. We further analyze the root causes of these misjudgments, and propose two improved prompting strategies for mitigation. For the first time, our findings reveals unrecognized limitations in LLMs to match code with requirements. We also offer novel insights and practical guidance for effective use of LLMs in automated code review and task-oriented agent scenarios.
\end{abstract}

%%
%% The code below is generated by the tool at http://dl.acm.org/ccs.cfm.
%% Please copy and paste the code instead of the example below.
%%



% \received{20 February 2007}
% \received[revised]{12 March 2009}
% \received[accepted]{5 June 2009}

\maketitle

\section{Introduction}
As large language models (LLMs) have demonstrated increasing capability in the domain of code synthesis \cite{austin2021program}, a growing body of research and tooling efforts have begun exploring their application for automated code review and verification \cite{xu2022systematic,rasheed2024ai,cai2025automated}. Traditional code reviews require developers to manually verify the alignment between code logic and requirements, a process that is both time consuming and prone to human error. LLMs show significant potential to reduce this burden by automating code assessments and suggesting improvements during the code-review process. For instance, recent studies have leveraged models such as GPT-4o to evaluate code submissions and determine whether they meet quality standards or require revisions \cite{liu2023your}.

% In software engineering, verifying and validating (V\&V) that source code aligns with its natural-language requirements remains an ongoing challenge \cite{shankar2024validates}. Requirements engineering is widely recognized as crucial to clearly defining, understanding, and aligning project objectives with stakeholder needs, thereby reducing risks associated with errors, delays, and project failures \cite{couder2024requirements}. Typically, verifying whether an implementation satisfies all requirements (and confirming the correctness of these requirements themselves) relies heavily on test cases or formal specifications, which are often unavailable or incomplete in real-world projects. Recent studies have begun to investigate the potential of LLMs to bridge this gap. Advanced models like GPT-4o have already shown promising accuracy, achieving F1-scores ranging from 79\% to 94\% when identifying unmet requirements from textual descriptions \cite{reinpold2024exploring}, indicating their potential as reliable "virtual reviewers" even without test cases or formal implementations. To further improve review effectiveness, researchers have explored self-critical mechanisms within LLMs, such as the Self-Refine framework, where models iteratively critique and refine their outputs without additional training \cite{madaan2023selfrefineiterativerefinementselffeedback}.

In software engineering, verifying and validating that source code aligns with its task requirements remains a challenge \cite{shankar2024validates}. Requirements engineering is widely recognized as crucial to clearly defining, understanding, and aligning project objectives with stakeholder needs, thereby reducing risks associated with errors, delays, and project failures \cite{couder2024requirements}. Recent studies start to investigate the potential of LLMs to bridge this gap. Advanced models such as GPT-4o have shown promising accuracy in identifying unmet requirements from textual descriptions \cite{reinpold2024exploring}, indicating their potential as reliable ``virtual reviewers'' even without test cases or formal implementations. To further improve review effectiveness, researchers have explored self-critical mechanisms within LLMs, such as the Self-Refine framework, where models iteratively critique and refine their outputs without additional training \cite{madaan2023selfrefineiterativerefinementselffeedback}.

Ideally, LLMs should accurately understand functional requirements described in natural language and reliably judge whether provided code satisfies these requirements, assisting developers in identifying defects or confirming correctness. However, in scenarios lacking test cases or reference implementations, the reliability of LLMs in performing such ``description-to-code'' evaluations remains unclear. In this work, we investigate whether LLMs can correctly determine code correctness when provided with precise task descriptions and correct implementations. Our preliminary experiments reveal a concerning phenomenon: LLMs frequently issue false negative judgments, incorrectly concluding that correct implementations fail to meet stated requirements. This systematic failure may arise from inherent biases and hallucinations within LLMs \cite{ji2023survey}, or from improper prompt designs. Models tend to overly critique correct code, resulting in a high false negative rate. Surprisingly, our extended experiments involving varied prompt designs \cite{white2023prompt} indicate that increasing the prompt complexity, such as requiring explicit explanations and suggested corrections, counterintuitively leads to higher rates of misjudgment. This finding contradicts the common assumption that incorporating explanatory steps typically enhances the reasoning and accuracy of LLMs. Instead, detailed prompts may inadvertently introduce biases toward excessive fault finding, causing models to detect non-existent errors in otherwise correct implementations.

In this paper, we present an empirical investigation with observations that carry significant implications for automated software engineering practices. Frequent false negatives by LLMs acting as automated code review assistants can severely impact their practical utility, overwhelming developers with misleading feedback and potentially diminishing trust in these tools. Particularly within automated code generation pipelines, where an LLM may evaluate its own generated code \cite{panickssery2024llm}, unreliable assessments will compromise the overall effectiveness of automated engineering workflow. Therefore, understanding why LLMs systematically misjudge correct implementations and identifying strategies to mitigate these failures becomes crucial. By providing these novel insights, we aim to raise community awareness regarding the limitations of LLMs in code evaluation and to inform future research efforts.
Overall, this work makes three key contributions:
\begin{enumerate}
    \item \textbf{False negatives discovery} – We reveal that LLMs frequently misjudge correct code as failing to meet requirements, indicating their bias towards over-correction rather than accurate verification.
    \item \textbf{Empirical evaluation} – Using the state-of-the-art models, our experiments highlight the key weaknesses in LLMs' ability to assess code conformance with natural language requirements.
    \item \textbf{Mitigation proposals with improved prompt strategies} – We propose a mitigation strategy that leverages refined prompt designs to reduce false negatives and improve assessment reliability.
\end{enumerate}
We will release the full replication package upon acceptance of this work. 
\begin{figure}
    \centering
    \includegraphics[width=\linewidth]{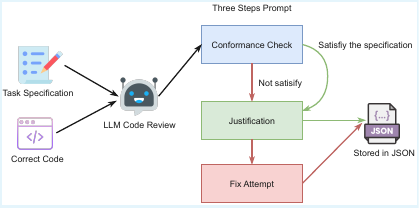}
    \caption{Experiment Workflow.}
    \label{fig:method}
\end{figure}

\section{Methodology}
\subsection{Research Question}
To investigate the causes and implications of LLMs' judgments in code evaluation, we propose the following research questions:
\begin{itemize}
    \item \textbf{RQ1:} Without test cases, to what extent can LLMs reliably assess whether a program conforms to its specification?
    \item \textbf{RQ2:} How does prompt design affect the accuracy of LLM-based code evaluation?
    \item \textbf{RQ3:} What factors cause LLMs to incorrectly classify correct code as faulty, and can these misjudgments be effectively mitigated through improved prompt design?
\end{itemize}
These research questions cover three key dimensions: understanding the root causes of LLMs' misjudgments, exploring potential mitigation solutions, and evaluating their effectiveness. Subsequent sections of this paper will present experiments and discussions addressing each of these questions.

\subsection{Preliminary Experiment Setup}
To address RQ1 and RQ2, we designed a series of experiments to evaluate the performance of different LLMs on code requirement conformance tasks and to analyze the impact of prompt design. For this purpose, we selected three widely used code evaluation benchmarks: HumanEval \cite{chen2021evaluating}, MBPP (Multiple Benchmark Programming Problems) \cite{austin2021programsynthesislargelanguage}, and QuixBugs \cite{ye2021comprehensive}, providing over 700 samples for evaluation in total. In this work, we currently focus on Python as the main programming language.

HumanEval provides manually written problem descriptions, corresponding reference implementations and test cases. We used the provided descriptions and correct reference implementations as model inputs, instructing models to judge if the implementation met the described requirements. MBPP includes programming tasks described in natural language along with correct implementations, and we similarly used these correct code implementations with their requirement descriptions for testing. QuixBugs contains a small set of algorithm problems with known defects. Through these diverse datasets, we covered various scenarios ranging from competitive programming solutions to real-world bug fixes, ensuring the generalizability of our observations.

Regarding model selection, we evaluated three mainstream LLMs: GPT-4o, Claude-3.5-sonnet, and Google Gemini-2.0 flash. These models were chosen as representatives of state-of-the-art industry LLMs, recognized for their strong general purpose capabilities in code understanding and generation tasks. Their superior performance makes them promising candidates for automated code review \cite{rasheed2024aipoweredcodereviewllms,fragiadakis2025evaluatinghumanaicollaborationreview,zhu2025judgelmfinetunedlargelanguage}. However, precisely because of their capabilities, we aimed to examine if they exhibit consistent error patterns. Moreover, these models have consistently been among the top choices for code review and code synthesis tasks in previous works \cite{10403378,joel2024surveyllmbasedcodegeneration,weyssow2024codeultrafeedbackllmasajudgedatasetaligning}.

Figure \ref{fig:method} illustrates the workflow: the LLM receives the task specification and corresponding correct code implementation, guided by our designed prompts, the LLM then assesses whether the code fulfills the stated requirements, initially performing a binary classification (``Yes'' if compliant, ``No'' if non-compliant). These judgments are systematically recorded into structured JSON files for consistency and ease of subsequent analysis. In cases where the LLM incorrectly classifies a correct implementation as non-compliant (a false negative judgment), it is further instructed to articulate a detailed rationale for its assessment and subsequently propose a revised version of the code. This approach enables the collection of enriched information for subsequent analysis of misjudgments.

% attempts to determine compliance, provides an explanation for its decision, and attempts repairs if non-compliance is identified. This approach enables the collection of enriched information for subsequent analysis of misjudgments.

\subsection{Prompt Design}
To ensure a fair comparison, we developed a unified three-step prompt template for all models which inspired by recent advances in prompt engineering \cite{wei2023chainofthoughtpromptingelicitsreasoning,yao2023reactsynergizingreasoningacting,jin2024graphchainofthoughtaugmentinglarge,araya2025chainsofthoughtslargelanguagemodels,akbar2024hallumeasure}, instructing each LLM model to complete the following tasks within a single round conversation:
\begin{itemize}
    \item \textbf{Judgment:} Read the natural language requirement and the provided code implementation, then answer the question, ``Does the code meet the requirement?'' with a simple answer ``Yes'' or ``No''.
    \item \textbf{Explanation:} Request the model to provide a rationale for the judgment, such as explaining why it believes the code does not meet the requirements. This covers a detailed analysis of any discrepancies between code logic and requirements.
    \item \textbf{Fix:} If the model judged the code as not meeting the requirement, it was instructed to provide corrected code after the explanation. If the code was deemed correct, this step could be skipped or explicitly noted as unnecessary.
\end{itemize}
% \begin{enumerate}
%     \item Read the natural language requirement and the provided code implementation, and initially answer, ``Does the code meet the requirement?'' with a simple ``Yes'' or ``No''.
%     \item Request the model to explain its reasoning, specifying why it believes the code does or does not meet the requirement, including a detailed analysis of discrepancies between the code logic and the requirement.
%     \item If the model judges the code as not meeting the requirement, instruct it to provide corrected code after its explanation. If the model determines the code to be correct, this step can be skipped or explicitly noted as unnecessary.
% \end{enumerate}
% \begin{figure}
%     \centering
%     \includegraphics[width=0.2\textwidth]{prompt.pdf}
%     \caption{Prompt Used In The Experiment.}
%     \label{fig:prompt}
% \end{figure}

Multi-stage prompting strategies have been applied for code evaluation task, aiming to enhance the reasoning performance of LLMs. One common approach first instructs the model to perform a step-by-step analysis of the code's functionality, and then asks it to summarize the analysis with a binary decision (correct or not). This approach outperforms simple prompts by encouraging LLMs to reason through both the requirements and the code logic before making a judgment~\cite{tong2024codejudge}. Such findings provide evidence-based support for our design of structured, explanation-driven prompting in code evaluation. Similar two-phase prompting method have also been applied for program repair. For example,~\cite{yin2024thinkrepair,wang2025mcts} prompt LLMs to first generate a chain-of-thought diagnosis of the bug, followed by providing a patch for fixing. Our prompt design is aligned with these approaches, in which the model is initially asked to explain any discrepancies between the code and the requirements. It will use related reasoning to propose a corrected version of the code if the initial answer is negative. 

%Outperformed simpler metrics by having the LLM reason through the requirements and code logic before deciding which directly inspires our approach of asking for an explanation after the initial judgment . Previous study also employs a two-phase prompt for program repair: first the model writes down a chain-of-thought diagnosing the bug, then it generates a patch to fix the error . Our prompt design follows a similar philosophy: the model initially explains any discrepancies between the code and requirements, then (if it answered ``No'') uses that reasoning to attempt a fix.

% We record the correctness of each model's judgement in relation to the accuracy of the known code, assessed the reasonableness and precision of the explanations provided (whether they correctly identified genuine problems or imagined problems) and verified if the proposed corrections genuinely satisfied the requirements. For incorrect code, we expected valid fixes whereas for originally correct code, unnecessary modifications indicated "over-fixing". Additionally, to study the impact of prompt complexity, we conducted experiments with prompt variations: simplifying the above three-step prompt by separately testing simpler prompts, one containing only a yes/no judgment directive and another two-step prompt not requiring code fixes. This allowed us to compare whether the misjudgment rate decreased when fewer details (explanations or corrections) were requested, thus analyzing the impact of individual prompt components on model decision making (addressing RQ2). 

\section{Observation and Preliminary Result}
\begin{table}
\setlength{\belowcaptionskip}{2pt}
\scriptsize
\centering
\caption{Requirement-Conformance Recognition Rate (RCRR ↑, in \%) of three LLMs on the HumanEval, MBPP, and QuixBugs benchmarks under three prompting approaches (Direct, Direct + Explain, and Three Step full prompt). Higher RCRR indicates better recognition of correct implementations as satisfying their task requirements.}
\label{tab:result}
\begin{tblr}{
  row{even} = {c},
  row{1} = {c},
  row{3} = {c},
  row{5} = {c},
  row{7} = {c},
  row{11} = {c},
  cell{1}{1} = {r=2}{},
  cell{1}{2} = {r=2}{},
  cell{3}{1} = {r=3}{},
  cell{5}{3} = {fg=red},
  cell{5}{4} = {fg=red},
  cell{5}{5} = {fg=red},
  cell{6}{1} = {r=3}{},
  cell{8}{3} = {fg=red},
  cell{8}{4} = {fg=red},
  cell{8}{5} = {fg=red},
  cell{9}{1} = {r=3}{},
  cell{9}{2} = {c},
  cell{9}{3} = {c},
  cell{9}{4} = {c},
  cell{9}{5} = {c},
  cell{11}{3} = {fg=red},
  cell{11}{4} = {fg=red},
  cell{11}{5} = {fg=red},
  hline{1,3,6,9,12} = {-}{},
  hline{2} = {3-5}{lr},
  hline{5,8,11} = {2-5}{dashed},
}
\textbf{Model}    & \textbf{Approch} & \textbf{HumanEval}       & \textbf{MBPP}            & \textbf{Quixbugs}        \\
                  &                  & \textbf{RCRR~$\uparrow$} & \textbf{RCRR~$\uparrow$} & \textbf{RCRR~$\uparrow$} \\
GPT-4o            & Direct           & 52.4                     & 63.7                     & 62.5                     \\
                  & Direct + Explain & 22.0                     & 38.2                     & 50.0                     \\
                  & Full             & 11.0~                    & 30.9                     & 30.0                     \\
Gemini-2.0-flash  & Direct           & 59.1                     & 66.5                     & 72.5                     \\
                  & Direct + Explain & 56.7                     & 64.9                     & 75.0                     \\
                  & Full             & 53.0                     & 61.5                     & 72.5                     \\
Claude-3-5-sonnet & Direct           & 78.0                     & 56.9                     & 80.0                     \\
                  & Direct + Explain & 69.5                     & 66.1                     & 77.5                     \\
                  & Full             & 67.0                     & 55.3                     & 75.0                     
\end{tblr}
\end{table}
To address our research questions, our experiments consider different dimensions for LLMs: accuracy of judgment against ground truth labels, and whether rationales identified real defects or hallucinations. To isolate the impact of prompt complexity, we designed three progressively more complex prompting strategies: (1) Direct (judgment only), (2) Direct+Explain (judgment with reasoning), and (3) Full (three-step prompt consisting of judgment, explanation, and suggested repair). We then compared the misjudgement rates across these conditions. Our evaluation metric was the Requirement-Conformance Recognition Rate (RCRR), representing the accuracy with which a model correctly judges whether a given code satisfies a natural-language requirement. The metric is formally defined in Equation \ref{equation}:

\begin{equation}
\label{equation}
\text{RCRR} = \frac{N_{\text{correct judgment}}}{N_{\text{Total correct\_code samples}}}
\end{equation}
Overall, our results indicate an inverse relationship between prompt complexity and model performance. For instance, on the HumanEval dataset, GPT-4o's RCRR declined markedly from 52.4\% with the simple direct prompt to merely 11.0\% using the full three-step prompt. It results in a substantial reduction of 41.4 percentage points. Similar trends were observed on the MBPP and QuixBugs datasets, with drops of 32.8 and 32.5 percentage points, respectively. Other models exhibited the same pattern: accuracy consistently decreased as explanation step and repair suggestions were introduced into the prompt. Notably, simpler ``judgment-only'' prompts often yielded the best performance, particularly on HumanEval and MBPP datasets. The consistency of these declines across diverse models and datasets underscores the robustness and generalizability of this phenomenon. 
% In summary, our preliminary results indicate that when LLMs are tasked with determining whether code meets specific requirements, adding explanations and repair steps akin to chain-of-thought reasoning \cite{wei2022chain} tends to harm rather than improve accuracy.

\section{Finding Evaluation}
Next, we examine the underlying reasons why multi-step prompts negatively impact accuracy. Our analysis identifies an increased rate of false negatives as the primary factor: when models are prompted to explain and propose fixes, models become significantly more prone to incorrectly classifying conforming code as erroneous. In other words, introducing reasoning and correction steps induces an overcorrection bias: LLMs tend to assume flaws exist and suggest unnecessary modifications, even when the provided implementation is already correct. This trend is particularly surprising, given contrasting findings from prior research that generally support step-by-step prompting to improve logical reasoning accuracy. For instance, Wei et al. demonstrate that intermediate reasoning steps significantly boost performance on complex reasoning tasks \cite{wei2022chain}. In contrast, a recent work also suggest similar bias may exist in the provided code snippet~\cite{moon2025don}. Our study extends this observation by establishing a more general scenario, in which we highlight a specific domain where the ``step-by-step'' approach can degrade rather than improve model performance in code requirement verification.

\subsection{\textbf{RQ1: LLM Judgment Accuracy}}
Regarding RQ1 (the capability of LLMs to judge code correctness without test cases), we find current LLMs exhibit limited effectiveness, with RCRR ranging from only \textbf{52\%} to \textbf{78\%}. This indicates that automated code-to-requirement conformance checking remains an unresolved challenge.

\subsection{\textbf{RQ2: Prompt Effectiveness}}
Addressing RQ2 (the impact of prompt design), we observe a significant effect of prompting strategy on model performance. In particular, the comprehensive three-step prompt (judgment, explanation, and fix) causes substantial accuracy reductions of approximately 20 to 40 percentage points compared to simpler direct prompts. GPT-4o exhibits the highest sensitivity to prompt complexity, with accuracy declines exceeding 40 percentage points on the most challenging tasks. In contrast, Gemini shows relatively greater stability, while Claude maintains comparatively better robustness across certain datasets. This variability indicates that different model architectures or training methodologies respond differently to chain-of-thought prompting strategies, though all exhibit some degree of performance degradation. Prior studies similarly report that even minor prompt modifications, such as slight variations in problem descriptions, can significantly affect LLM performance on code tasks \cite{kojima2022large}.

\definecolor{JapaneseLaurel}{rgb}{0,0.501,0}
\begin{table}
\scriptsize
\centering
\caption{RCRR for LLMs on the HumanEval, MBPP, and QuixBugs benchmarks under three prompting strategies. Original for the direct prompt and the Two-Phase Reflective and Behavioral Comparison prompts markedly outperform the Original prompt across all models and datasets.}
\label{tab:mitigation_result}
\begin{tblr}{
  row{1} = {c},
  row{2} = {c},
  row{3} = {c},
  row{4} = {c},
  row{6} = {c},
  row{7} = {c},
  row{10} = {c},
  cell{1}{1} = {r=2}{},
  cell{1}{2} = {r=2}{},
  cell{3}{1} = {r=3}{},
  cell{5}{3} = {c,fg=JapaneseLaurel},
  cell{5}{4} = {c,fg=JapaneseLaurel},
  cell{5}{5} = {c,fg=JapaneseLaurel},
  cell{6}{1} = {r=3}{},
  cell{7}{4} = {fg=JapaneseLaurel},
  cell{8}{3} = {c,fg=JapaneseLaurel},
  cell{8}{4} = {c},
  cell{8}{5} = {c,fg=JapaneseLaurel},
  cell{9}{1} = {r=3}{},
  cell{9}{2} = {c},
  cell{9}{3} = {c},
  cell{9}{4} = {c},
  cell{9}{5} = {c},
  cell{10}{3} = {fg=JapaneseLaurel},
  cell{10}{4} = {fg=JapaneseLaurel},
  cell{11}{3} = {c},
  cell{11}{4} = {c},
  cell{11}{5} = {c,fg=JapaneseLaurel},
  hline{1,3,6,9,12} = {-}{},
  hline{2} = {3-5}{lr},
  hline{4,7,10} = {2-5}{dashed},
}
\textbf{Model}    & \textbf{Approch}      & \textbf{HumanEval}       & \textbf{MBPP}            & \textbf{Quixbugs}        \\
                  &                       & \textbf{RCRR~$\uparrow$} & \textbf{RCRR~$\uparrow$} & \textbf{RCRR~$\uparrow$} \\
GPT-4o            & Original              & 52.4                     & 63.7                     & 62.5                     \\
                  & Two-Phase Reflective  & 72.0                     & 48.3                     & 70.0                     \\
                  & Behavioral Comparison & \textbf{85.4}            & \textbf{68.9}            & \textbf{90.0}            \\
Gemini-2.0-flash  & Original              & 59.1                     & 66.5                     & 72.5                     \\
                  & Two-Phase Reflective  & 67.7                     & \textbf{69.1}            & 82.5                     \\
                  & Behavioral Comparison & \textbf{73.8}            & 62.5                     & 85.0                     \\
Claude-3-5-sonnet & Original              & 78.0                     & 56.9                     & 80.0                     \\
                  & Two-Phase Reflective  & \textbf{82.9}            & \textbf{66.1}            & 77.5                     \\
                  & Behavioral Comparison & 78.0                     & 62.7                     & \textbf{82.5}            
\end{tblr}
\end{table}
\subsection{\textbf{RQ3: Analysis and Mitigation}}
Our error analysis identifies a pronounced ``over-correction'' tendency as the primary factor driving performance decline, when LLMs are prompted to explain and fix code, the model is required to justify its answer and then patch the code, it develops a bias toward assuming defects exist even when the implementation is already correct and the code already satisfies requirements, resulting in unnecessary and incorrect modifications. To address this issue, we developed two alternative prompting strategies, the Two-Phase Reflective Prompt and the Behavioral Comparison Prompt, which eliminate the mandatory ``fix'' step and clearly separate requirement comprehension from code auditing, thus reducing the tendency toward premature fault finding.

\begin{tcolorbox}
\textbf{Two-Phase Reflective Prompt}

``Phase 1 – Extract Contract Obligations.''  
Read the requirement and extract its intended functional obligations.  
List the main things the code is expected to do, including input-output behavior, edge-case handling, and any conditions or constraints.  

``Phase 2 – Audit and Verdict.''  
Carefully examine the code and check whether it fulfills each obligation you listed above.  
If an obligation is fully met, mark it as \textit{Satisfied}; if it is partially or incorrectly implemented, mark it as \textit{Not satisfied}.  
Based on the audit, decide:  
Does the code fulfill all essential obligations from the requirement?  
\end{tcolorbox}
\subsubsection{Two-Phase Reflective Prompt} Draws inspiration from contract based reviews in software engineering \cite{meyer1992applying}. It is structured in two stages: initially, the model explicitly extracts functional obligations from natural language requirements, presenting them clearly as bullet pointed lists to avoid over interpretation. In the second stage, the model audits the code implementation against each previously extracted obligation. This structured approach encourages the model to focus on actual functional requirements rather than superficial implementation details or stylistic differences, thereby effectively reducing the over correction bias. 

\subsubsection{Behavioral Comparison Prompt} Inspired by traditional black-box software testing, separately summarizes the core functionality, input-output behaviors, and boundary conditions from the requirement description and then independently summarizes the functionality actually implemented in the code. Finally, these two behavior descriptions are explicitly compared to determine consistency.
\begin{tcolorbox}
\textbf{Behavioral Comparison Prompt}

Please summarize the main functions and boundary conditions that the program should implement.  
Then read the code and describe what functions the code actually completes and how the key steps are implemented.  
Finally, compare the code behavior with the requirements point-by-point to determine whether they are consistent.  
\end{tcolorbox}

Table~\ref{tab:mitigation_result} illustrates the effectiveness of these new prompts. Experimental results show significant improvements in RCRR across models using the Two-Phase Reflective Prompt. For instance, GPT-4o's performance on HumanEval increased from 52.4\% with the original prompt to 72.0\%, an improvement of 61 percentage points, with Behavioral Comparison further elevating performance to 85.4\%. Similar trends are evident in the MBPP and QuixBugs datasets, highlighting the profound influence of prompt strategies on model performance. We attribute the superior effectiveness of these two prompting methods to their explicit redirection of the model's attention toward ``essential functional differences between requirements and code'', thus avoiding the overly critical stance commonly triggered by step-by-step explanation and correction prompts. This overly critical stance contradicts the original intent of classical Prompt Engineering and Chain-of-Thought methods \cite{wei2022chain, kojima2022large}, which aim to improve reasoning by introducing intermediate reasoning.

% Moreover, to rule out the possibility that models were overly cautious due to capability limitations, we constructed an adversarial dataset of faulty codes by manually injecting five types of bugs (logic errors, semantic algorithm errors, runtime crashes, output format errors, and performance bottlenecks) into 50 problems each from HumanEval and MBPP. Experiments showed GPT-4o detected 96\% of the bugs, Gemini detected 90\%, and Claude 80\%. These results clearly demonstrate that LLMs possess substantial capability to identify genuine functional errors. Consequently, the previously observed high misjudgment rates induced by earlier prompts were clearly not due to insufficient capability, but rather cognitive biases triggered by the prompting mechanisms.

\section{Conclusion}
In this paper, we have identified systematic misjudgments by LLMs when verifying code conformance to natural language task specifications. Our results revealed that increasing prompt complexity substantially reduces the models' Requirement Conformance Recognition Rate (RCRR), frequently leading to correct implementations being misclassified as non-conforming. 
This finding challenges prevailing prompt engineering assumptions, highlighting a critical limitation due to an ``over-correction'' bias. To address this issue, we have further proposed two prompt strategies for mitigation: the Two-Phase Reflective Prompt and the Behavioral Comparison Prompt. Experimental results demonstrated that both strategies improved LLM performance on code requirement verification tasks. While some misjudgments may persisted, we believe our findings offer new and promising directions for developing more reliable and effective automated code review tools and systems.

% \clearpage
\balance
\bibliographystyle{ACM-Reference-Format}
\bibliography{software}

%%% -*-BibTeX-*-
%%% Do NOT edit. File created by BibTeX with style
%%% ACM-Reference-Format-Journals [18-Jan-2012].

\begin{thebibliography}{33}

%%% ====================================================================
%%% NOTE TO THE USER: you can override these defaults by providing
%%% customized versions of any of these macros before the \bibliography
%%% command.  Each of them MUST provide its own final punctuation,
%%% except for \shownote{} and \showURL{}.  The latter two
%%% do not use final punctuation, in order to avoid confusing it with
%%% the Web address.
%%%
%%% To suppress output of a particular field, define its macro to expand
%%% to an empty string, or better, \unskip, like this:
%%%
%%% \newcommand{\showURL}[1]{\unskip}   % LaTeX syntax
%%%
%%% \def \showURL #1{\unskip}           % plain TeX syntax
%%%
%%% ====================================================================

\ifx \showCODEN    \undefined \def \showCODEN     #1{\unskip}     \fi
\ifx \showISBNx    \undefined \def \showISBNx     #1{\unskip}     \fi
\ifx \showISBNxiii \undefined \def \showISBNxiii  #1{\unskip}     \fi
\ifx \showISSN     \undefined \def \showISSN      #1{\unskip}     \fi
\ifx \showLCCN     \undefined \def \showLCCN      #1{\unskip}     \fi
\ifx \shownote     \undefined \def \shownote      #1{#1}          \fi
\ifx \showarticletitle \undefined \def \showarticletitle #1{#1}   \fi
\ifx \showURL      \undefined \def \showURL       {\relax}        \fi
% The following commands are used for tagged output and should be
% invisible to TeX
\providecommand\bibfield[2]{#2}
\providecommand\bibinfo[2]{#2}
\providecommand\natexlab[1]{#1}
\providecommand\showeprint[2][]{arXiv:#2}

\bibitem[Akbar et~al\mbox{.}(2024)]%
        {akbar2024hallumeasure}
\bibfield{author}{\bibinfo{person}{Shayan~Ali Akbar}, \bibinfo{person}{Md~Mosharaf Hossain}, \bibinfo{person}{Tess Wood}, \bibinfo{person}{Si-Chi Chin}, \bibinfo{person}{Erica~M Salinas}, \bibinfo{person}{Victor Alvarez}, {and} \bibinfo{person}{Erwin Cornejo}.} \bibinfo{year}{2024}\natexlab{}.
\newblock \showarticletitle{HalluMeasure: Fine-grained hallucination measurement using chain-of-thought reasoning}. In \bibinfo{booktitle}{\emph{Proceedings of the 2024 Conference on Empirical Methods in Natural Language Processing}}. \bibinfo{pages}{15020--15037}.
\newblock


\bibitem[Araya(2025)]%
        {araya2025chainsofthoughtslargelanguagemodels}
\bibfield{author}{\bibinfo{person}{Roberto Araya}.} \bibinfo{year}{2025}\natexlab{}.
\newblock \bibinfo{title}{Do Chains-of-Thoughts of Large Language Models Suffer from Hallucinations, Cognitive Biases, or Phobias in Bayesian Reasoning?}
\newblock
\showeprint[arxiv]{2503.15268}~[cs.AI]
\urldef\tempurl%
\url{https://arxiv.org/abs/2503.15268}
\showURL{%
\tempurl}


\bibitem[Austin et~al\mbox{.}(2021b)]%
        {austin2021program}
\bibfield{author}{\bibinfo{person}{Jacob Austin}, \bibinfo{person}{Augustus Odena}, \bibinfo{person}{Maxwell Nye}, \bibinfo{person}{Maarten Bosma}, \bibinfo{person}{Henryk Michalewski}, \bibinfo{person}{David Dohan}, \bibinfo{person}{Ellen Jiang}, \bibinfo{person}{Carrie Cai}, \bibinfo{person}{Michael Terry}, \bibinfo{person}{Quoc Le}, {et~al\mbox{.}}} \bibinfo{year}{2021}\natexlab{b}.
\newblock \showarticletitle{Program synthesis with large language models}.
\newblock \bibinfo{journal}{\emph{arXiv preprint arXiv:2108.07732}} (\bibinfo{year}{2021}).
\newblock


\bibitem[Austin et~al\mbox{.}(2021a)]%
        {austin2021programsynthesislargelanguage}
\bibfield{author}{\bibinfo{person}{Jacob Austin}, \bibinfo{person}{Augustus Odena}, \bibinfo{person}{Maxwell Nye}, \bibinfo{person}{Maarten Bosma}, \bibinfo{person}{Henryk Michalewski}, \bibinfo{person}{David Dohan}, \bibinfo{person}{Ellen Jiang}, \bibinfo{person}{Carrie Cai}, \bibinfo{person}{Michael Terry}, \bibinfo{person}{Quoc Le}, {and} \bibinfo{person}{Charles Sutton}.} \bibinfo{year}{2021}\natexlab{a}.
\newblock \bibinfo{title}{Program Synthesis with Large Language Models}.
\newblock
\showeprint[arxiv]{2108.07732}~[cs.PL]
\urldef\tempurl%
\url{https://arxiv.org/abs/2108.07732}
\showURL{%
\tempurl}


\bibitem[Cai et~al\mbox{.}(2025)]%
        {cai2025automated}
\bibfield{author}{\bibinfo{person}{Yufan Cai}, \bibinfo{person}{Zhe Hou}, \bibinfo{person}{David Sanan}, \bibinfo{person}{Xiaokun Luan}, \bibinfo{person}{Yun Lin}, \bibinfo{person}{Jun Sun}, {and} \bibinfo{person}{Jin~Song Dong}.} \bibinfo{year}{2025}\natexlab{}.
\newblock \showarticletitle{Automated Program Refinement: Guide and Verify Code Large Language Model with Refinement Calculus}.
\newblock \bibinfo{journal}{\emph{Proceedings of the ACM on Programming Languages}} \bibinfo{volume}{9}, \bibinfo{number}{POPL} (\bibinfo{year}{2025}), \bibinfo{pages}{2057--2089}.
\newblock


\bibitem[Chen et~al\mbox{.}(2021)]%
        {chen2021evaluating}
\bibfield{author}{\bibinfo{person}{Mark Chen}, \bibinfo{person}{Jerry Tworek}, \bibinfo{person}{Heewoo Jun}, \bibinfo{person}{Qiming Yuan}, \bibinfo{person}{Henrique Ponde De~Oliveira Pinto}, \bibinfo{person}{Jared Kaplan}, \bibinfo{person}{Harri Edwards}, \bibinfo{person}{Yuri Burda}, \bibinfo{person}{Nicholas Joseph}, \bibinfo{person}{Greg Brockman}, {et~al\mbox{.}}} \bibinfo{year}{2021}\natexlab{}.
\newblock \showarticletitle{Evaluating large language models trained on code}.
\newblock \bibinfo{journal}{\emph{arXiv preprint arXiv:2107.03374}} (\bibinfo{year}{2021}).
\newblock


\bibitem[Couder et~al\mbox{.}(2024)]%
        {couder2024requirements}
\bibfield{author}{\bibinfo{person}{Juan~Ortiz Couder}, \bibinfo{person}{Dawson Gomez}, {and} \bibinfo{person}{Omar Ochoa}.} \bibinfo{year}{2024}\natexlab{}.
\newblock \showarticletitle{Requirements verification through the analysis of source code by large language models}. In \bibinfo{booktitle}{\emph{SoutheastCon 2024}}. IEEE, \bibinfo{pages}{75--80}.
\newblock


\bibitem[Fragiadakis et~al\mbox{.}(2025)]%
        {fragiadakis2025evaluatinghumanaicollaborationreview}
\bibfield{author}{\bibinfo{person}{George Fragiadakis}, \bibinfo{person}{Christos Diou}, \bibinfo{person}{George Kousiouris}, {and} \bibinfo{person}{Mara Nikolaidou}.} \bibinfo{year}{2025}\natexlab{}.
\newblock \bibinfo{title}{Evaluating Human-AI Collaboration: A Review and Methodological Framework}.
\newblock
\showeprint[arxiv]{2407.19098}~[cs.HC]
\urldef\tempurl%
\url{https://arxiv.org/abs/2407.19098}
\showURL{%
\tempurl}


\bibitem[Ji et~al\mbox{.}(2023)]%
        {ji2023survey}
\bibfield{author}{\bibinfo{person}{Ziwei Ji}, \bibinfo{person}{Nayeon Lee}, \bibinfo{person}{Rita Frieske}, \bibinfo{person}{Tiezheng Yu}, \bibinfo{person}{Dan Su}, \bibinfo{person}{Yan Xu}, \bibinfo{person}{Etsuko Ishii}, \bibinfo{person}{Ye~Jin Bang}, \bibinfo{person}{Andrea Madotto}, {and} \bibinfo{person}{Pascale Fung}.} \bibinfo{year}{2023}\natexlab{}.
\newblock \showarticletitle{Survey of hallucination in natural language generation}.
\newblock \bibinfo{journal}{\emph{Comput. Surveys}} \bibinfo{volume}{55}, \bibinfo{number}{12} (\bibinfo{year}{2023}), \bibinfo{pages}{1--38}.
\newblock


\bibitem[Jin et~al\mbox{.}(2024)]%
        {jin2024graphchainofthoughtaugmentinglarge}
\bibfield{author}{\bibinfo{person}{Bowen Jin}, \bibinfo{person}{Chulin Xie}, \bibinfo{person}{Jiawei Zhang}, \bibinfo{person}{Kashob~Kumar Roy}, \bibinfo{person}{Yu Zhang}, \bibinfo{person}{Zheng Li}, \bibinfo{person}{Ruirui Li}, \bibinfo{person}{Xianfeng Tang}, \bibinfo{person}{Suhang Wang}, \bibinfo{person}{Yu Meng}, {and} \bibinfo{person}{Jiawei Han}.} \bibinfo{year}{2024}\natexlab{}.
\newblock \bibinfo{title}{Graph Chain-of-Thought: Augmenting Large Language Models by Reasoning on Graphs}.
\newblock
\showeprint[arxiv]{2404.07103}~[cs.CL]
\urldef\tempurl%
\url{https://arxiv.org/abs/2404.07103}
\showURL{%
\tempurl}


\bibitem[Joel et~al\mbox{.}(2024)]%
        {joel2024surveyllmbasedcodegeneration}
\bibfield{author}{\bibinfo{person}{Sathvik Joel}, \bibinfo{person}{Jie~JW Wu}, {and} \bibinfo{person}{Fatemeh~H. Fard}.} \bibinfo{year}{2024}\natexlab{}.
\newblock \bibinfo{title}{A Survey on LLM-based Code Generation for Low-Resource and Domain-Specific Programming Languages}.
\newblock
\showeprint[arxiv]{2410.03981}~[cs.SE]
\urldef\tempurl%
\url{https://arxiv.org/abs/2410.03981}
\showURL{%
\tempurl}


\bibitem[Kojima et~al\mbox{.}(2022)]%
        {kojima2022large}
\bibfield{author}{\bibinfo{person}{Takeshi Kojima}, \bibinfo{person}{Shixiang~Shane Gu}, \bibinfo{person}{Machel Reid}, \bibinfo{person}{Yutaka Matsuo}, {and} \bibinfo{person}{Yusuke Iwasawa}.} \bibinfo{year}{2022}\natexlab{}.
\newblock \showarticletitle{Large language models are zero-shot reasoners}.
\newblock \bibinfo{journal}{\emph{Advances in neural information processing systems}}  \bibinfo{volume}{35} (\bibinfo{year}{2022}), \bibinfo{pages}{22199--22213}.
\newblock


\bibitem[Liu et~al\mbox{.}(2023)]%
        {liu2023your}
\bibfield{author}{\bibinfo{person}{Jiawei Liu}, \bibinfo{person}{Chunqiu~Steven Xia}, \bibinfo{person}{Yuyao Wang}, {and} \bibinfo{person}{Lingming Zhang}.} \bibinfo{year}{2023}\natexlab{}.
\newblock \showarticletitle{Is your code generated by chatgpt really correct? rigorous evaluation of large language models for code generation}.
\newblock \bibinfo{journal}{\emph{Advances in Neural Information Processing Systems}}  \bibinfo{volume}{36} (\bibinfo{year}{2023}), \bibinfo{pages}{21558--21572}.
\newblock


\bibitem[Madaan et~al\mbox{.}(2023)]%
        {madaan2023selfrefineiterativerefinementselffeedback}
\bibfield{author}{\bibinfo{person}{Aman Madaan}, \bibinfo{person}{Niket Tandon}, \bibinfo{person}{Prakhar Gupta}, \bibinfo{person}{Skyler Hallinan}, \bibinfo{person}{Luyu Gao}, \bibinfo{person}{Sarah Wiegreffe}, \bibinfo{person}{Uri Alon}, \bibinfo{person}{Nouha Dziri}, \bibinfo{person}{Shrimai Prabhumoye}, \bibinfo{person}{Yiming Yang}, \bibinfo{person}{Shashank Gupta}, \bibinfo{person}{Bodhisattwa~Prasad Majumder}, \bibinfo{person}{Katherine Hermann}, \bibinfo{person}{Sean Welleck}, \bibinfo{person}{Amir Yazdanbakhsh}, {and} \bibinfo{person}{Peter Clark}.} \bibinfo{year}{2023}\natexlab{}.
\newblock \bibinfo{title}{Self-Refine: Iterative Refinement with Self-Feedback}.
\newblock
\showeprint[arxiv]{2303.17651}~[cs.CL]
\urldef\tempurl%
\url{https://arxiv.org/abs/2303.17651}
\showURL{%
\tempurl}


\bibitem[Meyer(1992)]%
        {meyer1992applying}
\bibfield{author}{\bibinfo{person}{Bertrand Meyer}.} \bibinfo{year}{1992}\natexlab{}.
\newblock \showarticletitle{Applying'design by contract'}.
\newblock \bibinfo{journal}{\emph{Computer}} \bibinfo{volume}{25}, \bibinfo{number}{10} (\bibinfo{year}{1992}), \bibinfo{pages}{40--51}.
\newblock


\bibitem[Moon et~al\mbox{.}(2025)]%
        {moon2025don}
\bibfield{author}{\bibinfo{person}{Jiwon Moon}, \bibinfo{person}{Yerin Hwang}, \bibinfo{person}{Dongryeol Lee}, \bibinfo{person}{Taegwan Kang}, \bibinfo{person}{Yongil Kim}, {and} \bibinfo{person}{Kyomin Jung}.} \bibinfo{year}{2025}\natexlab{}.
\newblock \showarticletitle{Don't Judge Code by Its Cover: Exploring Biases in LLM Judges for Code Evaluation}.
\newblock \bibinfo{journal}{\emph{arXiv preprint arXiv:2505.16222}} (\bibinfo{year}{2025}).
\newblock


\bibitem[Panickssery et~al\mbox{.}(2024)]%
        {panickssery2024llm}
\bibfield{author}{\bibinfo{person}{Arjun Panickssery}, \bibinfo{person}{Samuel Bowman}, {and} \bibinfo{person}{Shi Feng}.} \bibinfo{year}{2024}\natexlab{}.
\newblock \showarticletitle{Llm evaluators recognize and favor their own generations}.
\newblock \bibinfo{journal}{\emph{Advances in Neural Information Processing Systems}}  \bibinfo{volume}{37} (\bibinfo{year}{2024}), \bibinfo{pages}{68772--68802}.
\newblock


\bibitem[Rasheed et~al\mbox{.}(2024a)]%
        {rasheed2024ai}
\bibfield{author}{\bibinfo{person}{Zeeshan Rasheed}, \bibinfo{person}{Malik~Abdul Sami}, \bibinfo{person}{Muhammad Waseem}, \bibinfo{person}{Kai-Kristian Kemell}, \bibinfo{person}{Xiaofeng Wang}, \bibinfo{person}{Anh Nguyen}, \bibinfo{person}{Kari Syst{\"a}}, {and} \bibinfo{person}{Pekka Abrahamsson}.} \bibinfo{year}{2024}\natexlab{a}.
\newblock \showarticletitle{Ai-powered code review with llms: Early results}.
\newblock \bibinfo{journal}{\emph{arXiv preprint arXiv:2404.18496}} (\bibinfo{year}{2024}).
\newblock


\bibitem[Rasheed et~al\mbox{.}(2024b)]%
        {rasheed2024aipoweredcodereviewllms}
\bibfield{author}{\bibinfo{person}{Zeeshan Rasheed}, \bibinfo{person}{Malik~Abdul Sami}, \bibinfo{person}{Muhammad Waseem}, \bibinfo{person}{Kai-Kristian Kemell}, \bibinfo{person}{Xiaofeng Wang}, \bibinfo{person}{Anh Nguyen}, \bibinfo{person}{Kari Systä}, {and} \bibinfo{person}{Pekka Abrahamsson}.} \bibinfo{year}{2024}\natexlab{b}.
\newblock \bibinfo{title}{AI-powered Code Review with LLMs: Early Results}.
\newblock
\showeprint[arxiv]{2404.18496}~[cs.SE]
\urldef\tempurl%
\url{https://arxiv.org/abs/2404.18496}
\showURL{%
\tempurl}


\bibitem[Reinpold et~al\mbox{.}(2024)]%
        {reinpold2024exploring}
\bibfield{author}{\bibinfo{person}{Lasse~M Reinpold}, \bibinfo{person}{Marvin Schieseck}, \bibinfo{person}{Lukas~P Wagner}, \bibinfo{person}{Felix Gehlhoff}, {and} \bibinfo{person}{Alexander Fay}.} \bibinfo{year}{2024}\natexlab{}.
\newblock \showarticletitle{Exploring LLMs for Verifying Technical System Specifications Against Requirements}.
\newblock \bibinfo{journal}{\emph{arXiv preprint arXiv:2411.11582}} (\bibinfo{year}{2024}).
\newblock


\bibitem[Shankar et~al\mbox{.}(2024)]%
        {shankar2024validates}
\bibfield{author}{\bibinfo{person}{Shreya Shankar}, \bibinfo{person}{JD Zamfirescu-Pereira}, \bibinfo{person}{Bj{\"o}rn Hartmann}, \bibinfo{person}{Aditya Parameswaran}, {and} \bibinfo{person}{Ian Arawjo}.} \bibinfo{year}{2024}\natexlab{}.
\newblock \showarticletitle{Who validates the validators? aligning llm-assisted evaluation of llm outputs with human preferences}. In \bibinfo{booktitle}{\emph{Proceedings of the 37th Annual ACM Symposium on User Interface Software and Technology}}. \bibinfo{pages}{1--14}.
\newblock


\bibitem[Tong and Zhang(2024)]%
        {tong2024codejudge}
\bibfield{author}{\bibinfo{person}{Weixi Tong} {and} \bibinfo{person}{Tianyi Zhang}.} \bibinfo{year}{2024}\natexlab{}.
\newblock \showarticletitle{CodeJudge: Evaluating Code Generation with Large Language Models}.
\newblock \bibinfo{journal}{\emph{arXiv preprint arXiv:2410.02184}} (\bibinfo{year}{2024}).
\newblock


\bibitem[Wang and Chen(2023)]%
        {10403378}
\bibfield{author}{\bibinfo{person}{Jianxun Wang} {and} \bibinfo{person}{Yixiang Chen}.} \bibinfo{year}{2023}\natexlab{}.
\newblock \showarticletitle{A Review on Code Generation with LLMs: Application and Evaluation}. In \bibinfo{booktitle}{\emph{2023 IEEE International Conference on Medical Artificial Intelligence (MedAI)}}. \bibinfo{pages}{284--289}.
\newblock
\href{https://doi.org/10.1109/MedAI59581.2023.00044}{doi:\nolinkurl{10.1109/MedAI59581.2023.00044}}


\bibitem[Wang et~al\mbox{.}(2025)]%
        {wang2025mcts}
\bibfield{author}{\bibinfo{person}{Yutong Wang}, \bibinfo{person}{Pengliang Ji}, \bibinfo{person}{Chaoqun Yang}, \bibinfo{person}{Kaixin Li}, \bibinfo{person}{Ming Hu}, \bibinfo{person}{Jiaoyang Li}, {and} \bibinfo{person}{Guillaume Sartoretti}.} \bibinfo{year}{2025}\natexlab{}.
\newblock \showarticletitle{Mcts-judge: Test-time scaling in llm-as-a-judge for code correctness evaluation}.
\newblock \bibinfo{journal}{\emph{arXiv preprint arXiv:2502.12468}} (\bibinfo{year}{2025}).
\newblock


\bibitem[Wei et~al\mbox{.}(2023)]%
        {wei2023chainofthoughtpromptingelicitsreasoning}
\bibfield{author}{\bibinfo{person}{Jason Wei}, \bibinfo{person}{Xuezhi Wang}, \bibinfo{person}{Dale Schuurmans}, \bibinfo{person}{Maarten Bosma}, \bibinfo{person}{Brian Ichter}, \bibinfo{person}{Fei Xia}, \bibinfo{person}{Ed Chi}, \bibinfo{person}{Quoc Le}, {and} \bibinfo{person}{Denny Zhou}.} \bibinfo{year}{2023}\natexlab{}.
\newblock \bibinfo{title}{Chain-of-Thought Prompting Elicits Reasoning in Large Language Models}.
\newblock
\showeprint[arxiv]{2201.11903}~[cs.CL]
\urldef\tempurl%
\url{https://arxiv.org/abs/2201.11903}
\showURL{%
\tempurl}


\bibitem[Wei et~al\mbox{.}(2022)]%
        {wei2022chain}
\bibfield{author}{\bibinfo{person}{Jason Wei}, \bibinfo{person}{Xuezhi Wang}, \bibinfo{person}{Dale Schuurmans}, \bibinfo{person}{Maarten Bosma}, \bibinfo{person}{Fei Xia}, \bibinfo{person}{Ed Chi}, \bibinfo{person}{Quoc~V Le}, \bibinfo{person}{Denny Zhou}, {et~al\mbox{.}}} \bibinfo{year}{2022}\natexlab{}.
\newblock \showarticletitle{Chain-of-thought prompting elicits reasoning in large language models}.
\newblock \bibinfo{journal}{\emph{Advances in neural information processing systems}}  \bibinfo{volume}{35} (\bibinfo{year}{2022}), \bibinfo{pages}{24824--24837}.
\newblock


\bibitem[Weyssow et~al\mbox{.}(2024)]%
        {weyssow2024codeultrafeedbackllmasajudgedatasetaligning}
\bibfield{author}{\bibinfo{person}{Martin Weyssow}, \bibinfo{person}{Aton Kamanda}, \bibinfo{person}{Xin Zhou}, {and} \bibinfo{person}{Houari Sahraoui}.} \bibinfo{year}{2024}\natexlab{}.
\newblock \bibinfo{title}{CodeUltraFeedback: An LLM-as-a-Judge Dataset for Aligning Large Language Models to Coding Preferences}.
\newblock
\showeprint[arxiv]{2403.09032}~[cs.SE]
\urldef\tempurl%
\url{https://arxiv.org/abs/2403.09032}
\showURL{%
\tempurl}


\bibitem[White et~al\mbox{.}(2023)]%
        {white2023prompt}
\bibfield{author}{\bibinfo{person}{Jules White}, \bibinfo{person}{Quchen Fu}, \bibinfo{person}{Sam Hays}, \bibinfo{person}{Michael Sandborn}, \bibinfo{person}{Carlos Olea}, \bibinfo{person}{Henry Gilbert}, \bibinfo{person}{Ashraf Elnashar}, \bibinfo{person}{Jesse Spencer-Smith}, {and} \bibinfo{person}{Douglas~C Schmidt}.} \bibinfo{year}{2023}\natexlab{}.
\newblock \showarticletitle{A prompt pattern catalog to enhance prompt engineering with chatgpt}.
\newblock \bibinfo{journal}{\emph{arXiv preprint arXiv:2302.11382}} (\bibinfo{year}{2023}).
\newblock


\bibitem[Xu et~al\mbox{.}(2022)]%
        {xu2022systematic}
\bibfield{author}{\bibinfo{person}{Frank~F Xu}, \bibinfo{person}{Uri Alon}, \bibinfo{person}{Graham Neubig}, {and} \bibinfo{person}{Vincent~Josua Hellendoorn}.} \bibinfo{year}{2022}\natexlab{}.
\newblock \showarticletitle{A systematic evaluation of large language models of code}. In \bibinfo{booktitle}{\emph{Proceedings of the 6th ACM SIGPLAN international symposium on machine programming}}. \bibinfo{pages}{1--10}.
\newblock


\bibitem[Yao et~al\mbox{.}(2023)]%
        {yao2023reactsynergizingreasoningacting}
\bibfield{author}{\bibinfo{person}{Shunyu Yao}, \bibinfo{person}{Jeffrey Zhao}, \bibinfo{person}{Dian Yu}, \bibinfo{person}{Nan Du}, \bibinfo{person}{Izhak Shafran}, \bibinfo{person}{Karthik Narasimhan}, {and} \bibinfo{person}{Yuan Cao}.} \bibinfo{year}{2023}\natexlab{}.
\newblock \bibinfo{title}{ReAct: Synergizing Reasoning and Acting in Language Models}.
\newblock
\showeprint[arxiv]{2210.03629}~[cs.CL]
\urldef\tempurl%
\url{https://arxiv.org/abs/2210.03629}
\showURL{%
\tempurl}


\bibitem[Ye et~al\mbox{.}(2021)]%
        {ye2021comprehensive}
\bibfield{author}{\bibinfo{person}{He Ye}, \bibinfo{person}{Matias Martinez}, \bibinfo{person}{Thomas Durieux}, {and} \bibinfo{person}{Martin Monperrus}.} \bibinfo{year}{2021}\natexlab{}.
\newblock \showarticletitle{A comprehensive study of automatic program repair on the QuixBugs benchmark}.
\newblock \bibinfo{journal}{\emph{Journal of Systems and Software}}  \bibinfo{volume}{171} (\bibinfo{year}{2021}), \bibinfo{pages}{110825}.
\newblock


\bibitem[Yin et~al\mbox{.}(2024)]%
        {yin2024thinkrepair}
\bibfield{author}{\bibinfo{person}{Xin Yin}, \bibinfo{person}{Chao Ni}, \bibinfo{person}{Shaohua Wang}, \bibinfo{person}{Zhenhao Li}, \bibinfo{person}{Limin Zeng}, {and} \bibinfo{person}{Xiaohu Yang}.} \bibinfo{year}{2024}\natexlab{}.
\newblock \showarticletitle{Thinkrepair: Self-directed automated program repair}. In \bibinfo{booktitle}{\emph{Proceedings of the 33rd ACM SIGSOFT International Symposium on Software Testing and Analysis}}. \bibinfo{pages}{1274--1286}.
\newblock


\bibitem[Zhu et~al\mbox{.}(2025)]%
        {zhu2025judgelmfinetunedlargelanguage}
\bibfield{author}{\bibinfo{person}{Lianghui Zhu}, \bibinfo{person}{Xinggang Wang}, {and} \bibinfo{person}{Xinlong Wang}.} \bibinfo{year}{2025}\natexlab{}.
\newblock \bibinfo{title}{JudgeLM: Fine-tuned Large Language Models are Scalable Judges}.
\newblock
\showeprint[arxiv]{2310.17631}~[cs.CL]
\urldef\tempurl%
\url{https://arxiv.org/abs/2310.17631}
\showURL{%
\tempurl}


\end{thebibliography}

\end{document}